\documentclass[twocolumn,showpacs,preprintnumbers,amsmath,amssymb, aps]{revtex4}

\usepackage{graphicx}
\usepackage{dcolumn}
\usepackage{bm}

\begin{document}

\preprint{APS/123-QED}

\title{Cold atom confinement in an all-optical dark ring trap}

\author{Spencer E. Olson}
\author{Matthew L. Terraciano}%
\author{Mark Bashkansky}%
\author{Fredrik K. Fatemi}%

\affiliation{Naval Research Laboratory, 4555 Overlook Ave. S.W.,
Washington, DC 20375}

\date{\today}

\begin{abstract} We demonstrate confinement of $^{85}$Rb atoms in
a dark, toroidal optical trap.  We use a spatial light modulator to
convert a single blue-detuned Gaussian laser beam to a superposition
of Laguerre-Gaussian modes that forms a ring-shaped intensity null
bounded harmonically in all directions.  We measure a 1/$e$
spin-relaxation lifetime of $\approx$1.5 seconds for a trap detuning
of 4.0~nm.  For smaller detunings, a time-dependent relaxation rate
is observed.  We use these relaxation rate measurements and imaging
diagnostics to optimize trap alignment in a programmable manner with
the modulator. The results are compared with numerical simulations.
\end{abstract}

\maketitle

\noindent Toroidal traps for cold atoms have recently been of
interest for both fundamental and applied research.  A toroidal
geometry can enable studies of phenomena in non-simply connected or
low dimensional topologies~\cite{helmerson, gupta, jain, bludov,
jackson, lesanovsky, fernholz, morizot, dutta, arnold, wu,
ruostekoski}, {\it e.g.} superfluid persistent circulation states of
Bose-Einstein condensates (BECs)~\cite{helmerson}.  A ring-shaped
atom waveguide may also be suitable for inertial
measurements~\cite{gustavson} and neutral atom storage~\cite{dutta,
arnold, wu, sauer}.

Several approaches for generating ring-shaped waveguides have been
proposed and implemented.  Magnetic fields have been used to create
large ring traps for possible use as atom storage rings or Sagnac
interferometry~\cite{gupta, sauer, arnold, wu}. Helmerson {\it{et
al.}}~\cite{helmerson} used a combination of magnetic and optical
fields to demonstrate persistent current flow of a BEC. Morizot
{\it{et al.}}~\cite{morizot} proposed ring traps formed from the
combination of an optical standing wave with rf-dressed atoms in a
magnetic trap.

All-optical approaches have also been considered for toroidal traps
~\cite{wright,courtade, freegarde}. Wright
{\it{et~al.}}~\cite{wright} suggested the use of
high-azimuthal-order Laguerre-Gaussian (LG) beams to confine atoms
with red-detuning. Atoms in red-detuned optical traps seek high
intensity, and with large detuning, spontaneous photon scattering
can be negligible. Photon scattering can also be reduced by using
blue-detuned optical traps.  Such ``dark'' traps confine atoms to
low intensity, allowing field-free measurements~\cite{ozeri, grimm,
friedman, kaplan}, but are challenging to make because they require
an intensity minimum bounded by higher intensity.  This challenge is
often overcome by crossing beams~\cite{kuga, fatemi_SLM, friedman}
to plug a hollow optical potential, although dark point atom traps
have been realized with a single laser beam containing a
phase-engineered intensity null~\cite{ozeri}.  The single beam
approach has the advantage of alignment simplicity over crossed-beam
configurations. Lattices of dark rings have been
proposed~\cite{freegarde} and realized~\cite{courtade} using
counterpropagating laser beams, but to the best of our knowledge,
there have been no reports of atom confinement in a lone optical
ring trap.

In this paper, we report atom confinement within a different class
of dark optical ring traps.  We form a bounded, ring-shaped
intensity null by converting a Gaussian laser beam to a dual-ringed
beam with a programmable spatial light modulator (SLM). SLMs are of
increasing value in cold atom manipulation experiments because of
their ability to reconfigure trap parameters~\cite{fatemi_SLM,
bergamini_SLM, chattrapiban_SLM, mcgloin_SLM, boyer_SLM}. We measure
the spin-relaxation lifetime, observe atom dynamics within the
traps, and compare the experimental results with numerical
simulations.

\begin{figure}
\includegraphics[width=8.4cm]{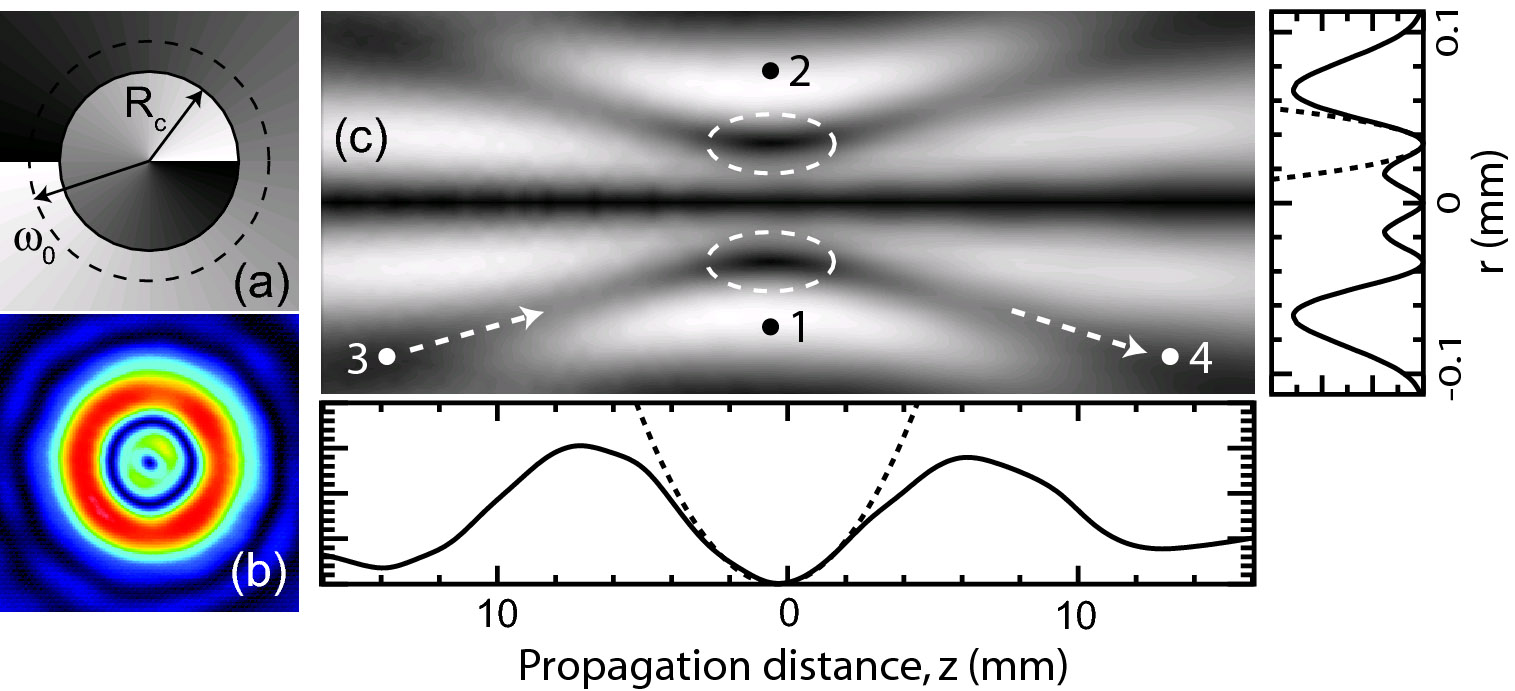}
\caption{\label{fig:beamprofiles} (Color online) a)  Phase profile
for creating the dark optical ring.  b) CCD image of a dual-ringed
beam in the focal plane. c) Numerical simulation of $r$-$z$ cross
section. Right: Transverse profile through the minimum, on a line
through points 1 and 2. Bottom: Profile along the minimum-intensity
path, indicated by the arrows, through points 3 and 4. Length scales
are shown on the profile plots.  The dashed curves in the profile
plots are quadratic fits. The dark trap is formed in a ring through
the centers of the dashed ellipses.}
\end{figure}

We form the dual-ringed laser beam by modifying the spatial phase of
a laser beam with an SLM, in a similar manner to that used for
producing hollow laser beams~\cite{fatemi_SLM, chattrapiban_SLM,
rhodes}. The latter can be created by imparting an azimuthal phase
$\Phi(r, \phi) = \ell\phi$, with integer $\ell$, to a Gaussian laser
beam $E(r)=|E_0|exp\left(-r^2/w_0^2\right)$, where $w_0$ is the
waist. The phase discontinuity at $r$ = 0 results in a hollow beam
that, for low $\ell$, closely approximates a pure $LG_{p=0}^\ell$
mode, where $p$ and $\ell$ are radial and azimuthal indices. As
shown in Figs.~\ref{fig:beamprofiles}a-b, a dual ring is produced by
introducing a $\pi$ phase discontinuity at $r = R_c > 0$ such that
the resulting beam has large overlap with the $LG_{p=1}^\ell$ mode,
which has two radial nodes. The parameter $R_c/w_0$ controls the
modal composition and thus the propagation characteristics. In
Ref.~\cite{arlt2}, $R_c/w_0$ was set to generate high purity LG
modes. Here, we adjust $R_c/w_0$ to create a superposition of
$LG_p^\ell$ modes that produces a dark ring at the focus of a lens
that is bounded in both the radial and longitudinal directions.

Figure~\ref{fig:beamprofiles}c shows the calculated $r$-$z$
cross-section of a toroidal beam with $\ell=1$ as it propagates
along $z$ through the focus of an $f$=215~mm focal length lens
($w_0=1.7$mm). We have chosen values of $R_c$ such that the barrier
heights in the longitudinal and transverse directions are equal. For
$\ell$=0, 1, and 2, this condition is satisfied for $R_c/w_0
\approx$ 0.71, 0.79, and 0.85. The small numerical aperture
(NA=$w_0/f$=0.008) leads to a long aspect ratio of $\approx$1:300
for $\ell=1$, defined as the ratio of the longitudinal trap
frequency $\omega_\parallel$ to the transverse trap frequency
$\omega_\perp$. The mode composition is dominated by $p=0$
(single-ringed) and $p=1$ (dual-ringed) modes.  For $\ell=0$,
\emph{e.g.}, the $p=0(1)$ fraction is 13\%(78\%).  The potential is
harmonic in all directions, as indicated in
Fig.~\ref{fig:beamprofiles}c. Under these conditions, the ratio of
the inner radial barrier height to the outer radial barrier height
is $\approx$25-35\%.  The radius of the trap depends linearly on
$\ell$, as it does for hollow beams~\cite{curtis, fatemi_ao}.

The trapping beam is derived from a 30~mW extended cavity diode
laser tunable from 776-780~nm. The beam is amplified to
$\approx$350~mW with a tapered amplifier of which $\approx$150~mW is
coupled into polarization maintaining fiber.  The linearly polarized
fiber output is collimated with $w_0$=1.7~mm, and reshaped by a
512x512 reflective SLM (Boulder Nonlinear Systems) with 15~$\mu$m
pixels and $\approx$90\% absolute diffraction efficiency.  A 4-$f$
imaging setup relays this modified Gaussian beam to a
magneto-optical trap (MOT).  The 4-$f$ relay roughly positions the
focus of the ring trap over the MOT, but fine longitudinal
adjustments are controlled entirely by the SLM by adding a lens
phase profile $\Phi_{\rm{lens}}(r, \phi) = -{\pi}r^2/f\lambda$.  We
compensate for wavefront errors imposed by the SLM by calibrating
the programmed phase on a pixel-by-pixel basis.

The experiment begins with a MOT containing $10^{7}$ $^{85}$Rb
atoms. After a 1 second loading time, the MOT coils are shut off,
and the atoms are cooled to 5~${\mu}K \approx\hbar\Gamma/60k_B$
during a 10~ms molasses cooling stage. All cooling and trapping
beams are then extinguished, followed by a 100~$\mu$s pulse that
optically pumps the atoms into the $F$=2 hyperfine level. The
toroidal beam power is ramped to $\approx$150~mW over 5~ms during
the molasses stage. This ramp adiabatically loads atoms into the
trap and minimizes the energy gained in the loading process. The
trap diameter is significantly smaller than the initial MOT size, so
we typically load only a small fraction of atoms
($\approx$5$\times$10$^4$) into the traps.  Collisions with
background gas limit the trap 1/$e$ lifetimes to $\approx$1~s. After
a variable delay, the trapped atoms are imaged onto an
electron-multiplying CCD camera (Andor Luca) by a 500~$\mu$s pulse
from the MOT and repump beams. Immediately prior to the imaging
pulse, the trapping beam is switched off to avoid Stark shifting of
the levels.  For linear polarization, the optical potential
is~\cite{grimm}

\begin{equation}
U(r) = \frac{\hbar{\Gamma}I(r)}{24I_s}\left(\frac{\Gamma}{\Delta +
\Delta_{\rm{LS}}} + \frac{2\Gamma}{\Delta}\right)
\end{equation}

\noindent where $I_{\rm s}$=1.6~mW/cm$^2$ is the saturation
intensity, $\Gamma$=2$\pi\times$6.1~MHz is the linewidth, and
$\Delta_{\rm{LS}}$=2$\pi\times$ 7.1~THz (=15~nm) is the fine
structure splitting. The resulting trap depths for $\ell=1$ and
$\Delta=0.5$ nm, 1.0 nm, 2.0 nm, and 4~nm are 0.26$\hbar\Gamma$,
0.13$\hbar\Gamma$, 0.065$\hbar\Gamma$, and 0.033$\hbar\Gamma$ (at
780~nm, 1~nm$\leftrightarrow$493~GHz). At $\Delta=1$~nm,
$\omega_\perp{\approx}2\pi\times$800~Hz and
$\omega_\parallel{\approx}2\pi\times$3~Hz.

\begin{figure}
\includegraphics[width=8.4cm]{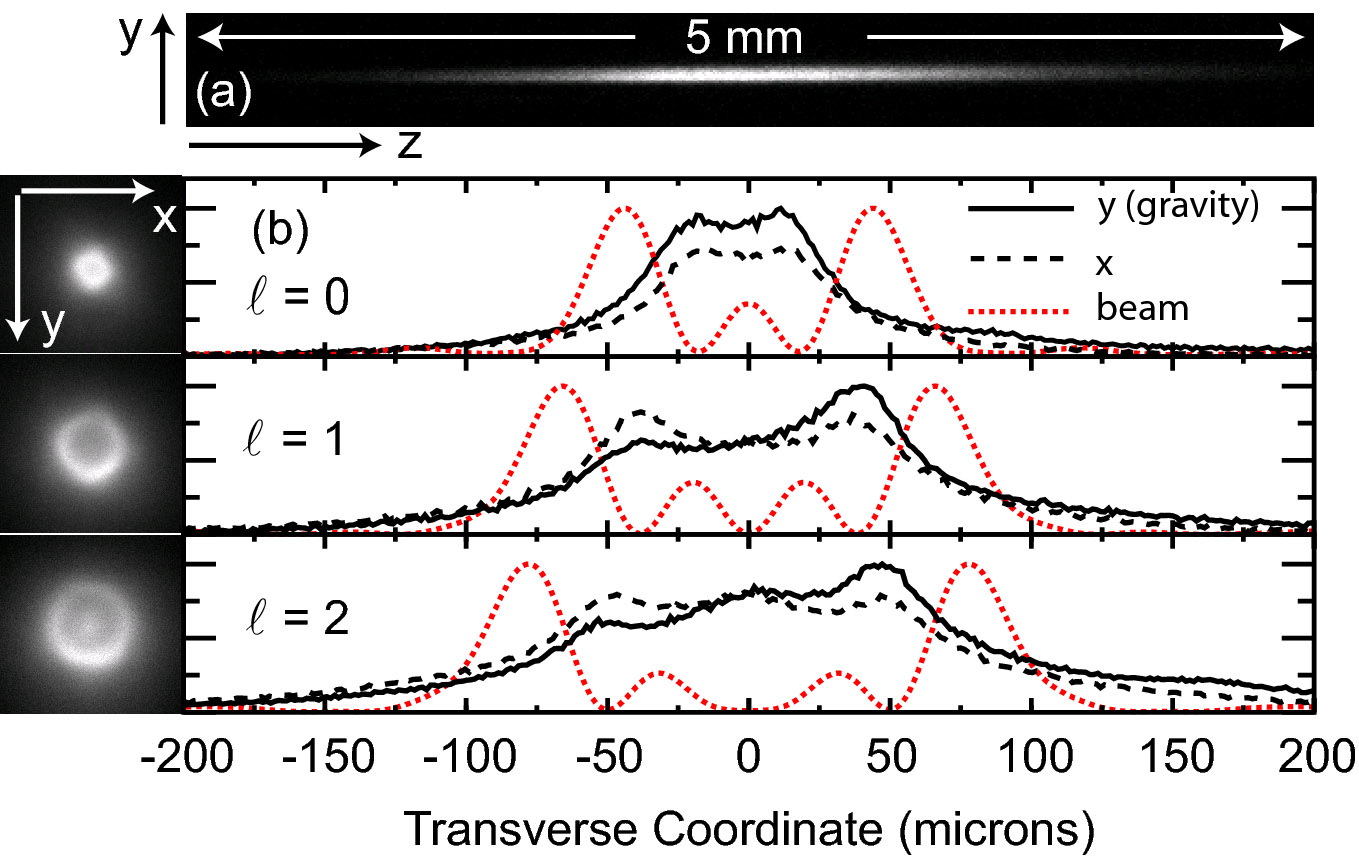}
\caption{\label{fig:atoms-and-beams} (Color online) a) Image of atom
cloud taken along $x$-axis for $\ell=1$. b) Images taken along
$z$-axis (left) and profiles (right) for toroidal traps using
$\ell=0-2$. The experimental beam profiles are shown for comparison
(dotted line). Trap time for this figure is 600 ms.}
\end{figure}

We record images of the trapped atoms with the camera axis along $x$
and along $z$.  Images along $x$ show the longitudinal trap extent
(Fig.~\ref{fig:atoms-and-beams}a), while those along $z$ show the
toroidal structure (Fig.~\ref{fig:atoms-and-beams}b). The head-on
views in Fig.~\ref{fig:atoms-and-beams}b are taken after a trap time
of 600~ms for $\ell = 0{\rm-}2$. Also shown are the
azimuthally-averaged beam intensity profiles in the focal plane and
atom distributions in the $x$ and
$y$ (gravity) directions.  
Because the trapping beam is propagating horizontally, the potential
is not azimuthally symmetric. The gravitational potential energy
difference between the intensity nulls for $\ell=2$ is $\Gamma$/30
$\approx 2\pi\times$200 kHz for $^{85}$Rb, which is larger than the
atom cloud temperature of 2$\pi\times$100~kHz. Thus, most atoms are
found in the bottom portion of the trap.  For $\ell\geq1$, atoms
could initially be loaded on the axis of the beam, along which there
is no barrier. This is seen for $\ell=2$ in
Fig.~\ref{fig:atoms-and-beams}.  In our configuration it takes a few
seconds for these atoms to drift away. Although there should be
little interaction between axial atoms and the ring-trapped atoms
under adiabatic loading, the axial atoms can be reduced by several
means, such as orienting the trapping beam vertically, or loading
from an atom distribution that has been dimpled by a blue-detuned
Gaussian beam, as in Ref.~\cite{helmerson}. A vertical propagation
axis would permit a symmetric ring potential in a horizontal plane,
but optical access in this direction was limited.

Imaging constraints prevent high contrast images of the toroidal
atom distributions. We use an 85~mm Nikon f/1.4 lens, the front
element of which is $\approx$250~mm away from the trap location.
This lens collects the maximum fluorescence and achieves a peak
resolution of $\approx$5~$\mu$m but suffers from spherical
aberration, which causes the observed loss of contrast.

\begin{figure}
\includegraphics[width=8.4cm]{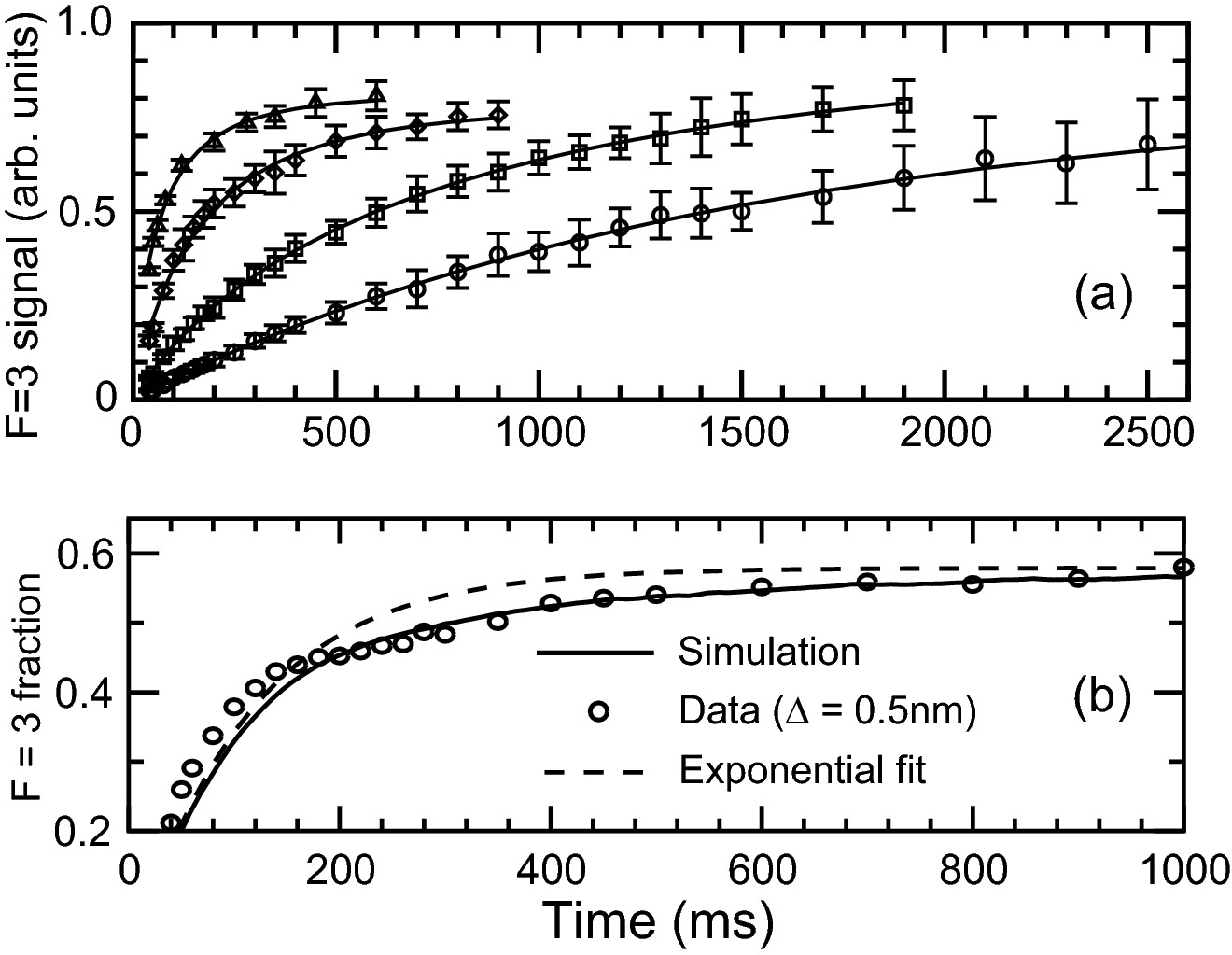}
\caption{\label{fig:state-lifetimes} a) Measurement of F=3 fraction
as a function of time for $\Delta$=0.5nm (triangles), 1.0 nm
(diamonds), 2.0 nm (squares), and 4 nm (circles).  Fits (solid
lines) using the model described in the text.  b) Comparison of
0.5~nm data with a single parameter exponential curve and
simulations.}
\end{figure}

One benefit of dark traps for coherent atom manipulation is the
suppression of photon scattering events~\cite{ozeri, grimm,
friedman, kaplan}. We measure the spin relaxation rate due to Raman
scattering by measuring the fraction of atoms in the trap that
transfer to $F$=3 as a function of trap~\cite{MillerPRA1993}. The
atoms are first pumped into the $F$=2 hyperfine level.  After a
variable trapping time, we image only the atoms that transfer to
$F$=3 by using a 10~$\mu$s pulse of resonant cycling transition
light. Within 2~ms, both the repump and the cycling transition beams
are switched on to image the atoms in both the $F$=2 and $F$=3
states. For background subtraction, two images with the same pulse
sequence are taken with no atoms present. This type of background
subtraction is necessary to eliminate false counts due to CCD
ghosting.  By taking the images during a single loading cycle, the
effect of atom number fluctuations is reduced. These images are
recorded along $x$ (as in Fig.~\ref{fig:atoms-and-beams}a). Between
the first two imaging pulses, the atom distribution expands slightly
beyond the few integrated rows of pixels. This leads to a slightly
low estimate of the total atom count, but the resulting $F$=3
normalized signal is proportional to the actual $F$=3 fraction.

We record the $F$=3 signal fraction as a function of trap time for
four different detunings (Fig.~\ref{fig:state-lifetimes}a).  In the
simplest approximation that all atoms have an equal scattering rate,
each curve can be modeled by a single exponential $N_3(t) = C(1$ -
exp(-${t/\tau}$)), as was used in Ref.~\cite{ozeri}, where $\tau$ is
the 1/$e$ decay time. For $\Delta \leq 1$~nm, however, a single
relaxation rate was not observed (Fig.~\ref{fig:state-lifetimes}b).
This difference between our results and those of Ref.~\cite{ozeri}
is most likely due to differences in the trap loading technique,
which we have found to affect the rate curves. We note that the
$F$=3 fraction at long times should approach 7/12, but our measured
values are higher due to the pixel integration described above.

Instead of modeling the spin-relaxation with a single-parameter time
constant, we phenomenologically ``chirp'' $\tau$ to be $\tau$(t) =
$\tau_0 + \beta{t^{1/2}}$ so that we can estimate the relaxation
rate at different times. We choose a sublinear chirp rate so that
the exponential will decay at long times, but the exact functional
form will depend on trap geometry.  A steadily increasing $\tau$
should be expected since atoms initially loaded into the trap in
locations of high intensity scatter photons more quickly than those
loaded into the dark portions of the trap. Thus, a rapid increase in
the $F$=3 fraction is observed for small $t$, followed by longer
relaxation times for the atoms with the least total energy. Using
this form for the $F$=3 fraction, approximate spin-relaxation
lifetimes at t = 0 for $\Delta=0.5$~nm, 1.0~nm, 2.0~nm, and $4.0$~nm
are 35~ms, 115~ms, 460~ms, and 1440~ms; after 500~ms, these increase
to 140~ms, 230~ms, 750~ms, and 1500~ms.

The scattering time for atoms in a red-detuned trap of comparable
depth at $\Delta$=0.5nm would be $\approx$2.5~ms, which is 50 times
shorter than our recorded value. In Ref.~\cite{ozeri}, the
blue-detuned trap had a scattering lifetime 700 times longer than a
comparable red-detuned trap at 0.5~nm. That work used significantly
higher intensities, where the differences between red- and
blue-detuning are more dramatic. Photon scattering may be reduced
substantially by using commercially available lasers with higher
power and larger detuning. For $\Delta > \Delta_{\rm{LS}}$, spin
relaxation is further suppressed, asymptotically scaling as
$\Delta^{-4}$~\cite{MillerPRA1993}. The time-dependent scattering
rate is likely not limited to toroidal geometries, but to the best
of our knowledge, it has been observed for the first time in this
report. Also, we point out that we did not directly measure the
recoil scattering rate, but for our $\Delta$ this is on the same
order as the spin-relaxation rate. A recoil scattering rate of
1~s$^{-1}$ corresponds to a heating rate of $\approx$400 nK/s.

To demonstrate the time-dependent scattering rate numerically, we
perform Monte Carlo simulations for $\Delta = 0.5$~nm.  The
simulated trap is ramped on over 5~ms.  The atom cloud is initially
in the $F =2$ state and normally distributed in position and
velocity to match the MOT size and temperature. Molasses effects are
ignored. Within each time step, each atom's hyperfine level is
changed with a probability determined by the local scattering rate,
as calculated from the Kramers-Heisenberg formula
~\cite{MillerPRA1993}. The simulation results, compared to data in
Fig.~\ref{fig:state-lifetimes}b, confirm the time-dependent
relaxation rate described above.  For comparison, the data have been
renormalized to have an asymptotic value of $7/12$.

\begin{figure}
\includegraphics[width=8.4cm]{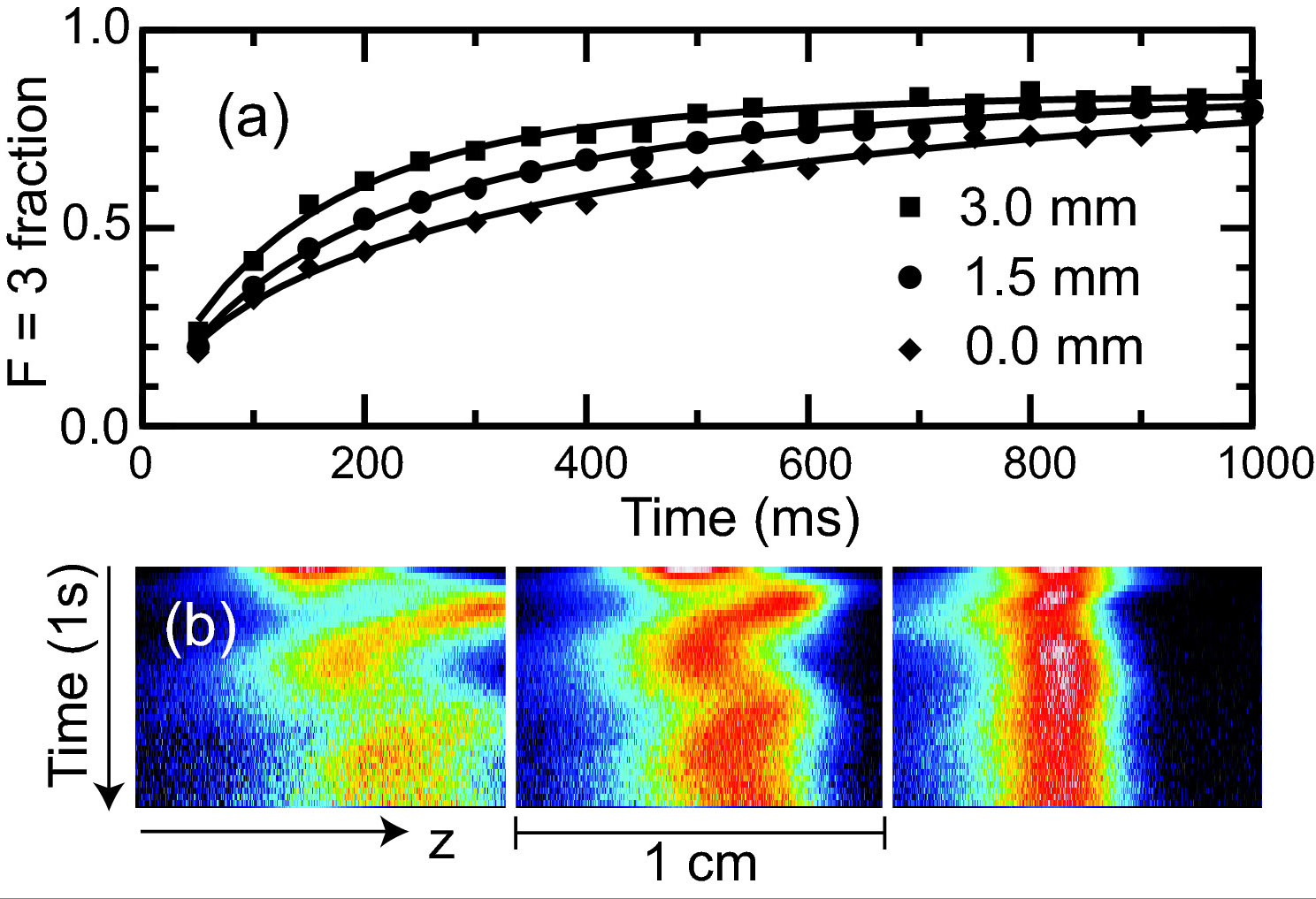}
\caption{\label{fig:sloshing} (Color online) a) Spin-relaxation
curves for three different starting positions. b) Images of
longitudinal oscillation of the atom cloud using trap displacement
from left to right of 3~mm, 1.5~mm, and 0.0mm.}
\end{figure}

To demonstrate axial confinement and to quantify the dependence of
the scattering rates on the starting position of the atoms in the
trap, we displace the trap minimum from the MOT by adjusting the
lens function written to the SLM by a few MOT radii (MOT radius
$\approx250~\mu$m).  Thus, most atoms are initially located in
regions of high intensity, reducing the overall scattering lifetime.
When the atom cloud is displaced 3~mm, 1.5~mm, and 0.0~mm away from
the trap minimum, the single-parameter rate constants (for
$\Delta=1$~nm) are 145~ms, 195~ms, and 230~ms
(Fig.~\ref{fig:sloshing}a). For each displacement, we show a
composite image of the side views of the trap
(Fig~\ref{fig:sloshing}b), where each row in the image is a
different slice in time. These images show the atom cloud
oscillating in the longitudinal direction when the trap is not well
overlapped with the atom cloud.  By displacing the trap focus, we
can also estimate the longitudinal trap frequency. For
$\Delta=1$~nm, we measure $\omega_\parallel\approx2\pi\times2$~Hz.
This agrees well with the estimate of
$\omega_\parallel\approx2\pi\times3$~Hz from the calculated
intensity profiles shown in Fig.~\ref{fig:beamprofiles}. The
scattering rate data and the composite images can be used to
optimize the location of the trap focus, which is done to
$\approx$100~$\mu$m with the SLM.

As with all single-beam traps, the aspect ratio scales with the
inverse of the trapping beam NA. For similar beam parameters, an
aspect ratio of $\approx$10 could be realized by using a $f=10$~mm
lens. A crossed beam geometry, in which additional beams cap the
potential in the longitudinal direction, allows significantly
tighter longitudinal confinement and larger diameter traps.  In
these cases, the ratio $R_c/w_0$ can be changed for optimal
confinement. One possibility is to use values of $R_c/w_0$ such that
the modified beam is primarily in a single $LG_{p=1}^{\ell}$
mode~\cite{arlt2}. Pure $LG_{p=1}^{\ell}$ modes have a radial
intensity null that persists for all values of $z$. When $R_c/w_0$
is chosen such that the most pure $LG_{p=1}^\ell$ is formed, the
inner radial barrier height is roughly 3$\times$ larger than the
outer one, and the longitudinal barrier is minimized.
 Therefore, the crossing beam can be well outside the focal plane,
where better beam quality is observed but the ring-shaped null
remains dark. The reduction of aberration effects outside the focal
plane was shown for hollow beams in Ref.~\cite{fatemi_ao}.

We have used a spatial light modulator to generate superpositions of
LG modes that form single-beam, dark ring traps for cold atoms. We
have shown that the atoms can be held in these potentials with long
state lifetimes.  We have observed atom dynamics in the longitudinal
direction and shown that by modifying the trap alignment with the
SLM we can optimize the scattering lifetime. This work was funded by
the Office of Naval Research and the Defense Advanced Research
Projects Agency.


\end{document}